\documentclass[11pt]{article}
\usepackage{psfig,epsf,graphics}

\date{\today}

\begin{document}
\begin{titlepage}
\begin{center}
{\large\bf Scale Invariance and Self-averaging in disordered systems.}\\[.3in]
  {\bf Giorgio Parisi$^{a}$, Marco Picco$^{b}$ and Nicolas Sourlas$^{c}$} \\
  $^{a}$ Dipartimento di Fisica, INFM, SMC and INFN,\\
  Universit\`a di Roma {\em La Sapienza},\\
  P. A. Moro 2, 00185 Rome, Italy\\
  $^{b}$ Laboratoire de Physique Th\'eorique et Hautes Energies
         \footnote{Unit{\'e} Mixte de Recherche CNRS UMR 7589, associ\'ee
         \`a l'Universit\'e  Pierre et Marie Curie, PARIS VI 
         et \`a l'Universit\'e  Denis Diderot, PARIS VII.},\\
  Bo\^{\i}te 126, Tour 16, 1$^{\it er}$ {\'e}tage,\\
  4 place Jussieu, F-75252 Paris CEDEX 05, France\\
  $^{c}$ Laboratoire de Physique Th\'eorique de l'Ecole Normale Sup\'erieure
         \footnote{Unit{\'e} Mixte de Recherche du CNRS et de
         l'Ecole Normale Sup\'erieure, associ\'ee
         \`a l'Universit\'e  Pierre et Marie Curie, PARIS VI.},\\
  24 rue Lhomond, 75231 Paris CEDEX 05, France

\end{center}
\vskip .15in
\centerline{\bf ABSTRACT}

\begin{quotation}

In a previous paper\cite{PS0} we found that in the random field Ising
model at zero temperature in three dimensions the correlation length
is not self-averaging near the critical point and that the violation
of self-averaging is maximal.  This is due to the formation of bound
states in the underlying field theory. We present a similar study for
the case of disordered Potts and Ising ferromagnets in two dimensions
near the critical temperature. In the random Potts model the
correlation length is not self-averaging near the critical temperature
but the violation of self-averaging is weaker than in the random field
case. In the random Ising model we find still weaker violations of
self-averaging and we cannot rule out the possibility of the
restoration of self-averaging in the infinite volume limit.



\end{quotation}
\end{titlepage}


One of the important issues in the physics of disordered systems is
the understanding of which physical quantities are self averaging,
i.e. for which observables the sample to sample fluctuations vanish
in the thermodynamic limit\cite{sou0,REF}.

In a recent work\cite{PS0} we studied the behavior of the correlation
length (CL) of the random field Ising model (RFIM) in three dimensions
at zero temperature. We found surprisingly that the CL is not
self-averaging and that the violation of self-averaging is maximal in
the vicinity of the critical point. This is due to the formation of
bound states in the underlying replica field theory. The interaction
among replicas is attractive and therefore it has the capability of
forming bound states. In the formation of bound states there is
competition between the strength of the attractive forces and the size
of the available phase space.  We have argued that in two dimensions
phase space is small and even a small attraction should win.

In this paper we address the question of the existence of bound states
and their consequence, the non self-averaging of the CL, in the case
of two disordered ferromagnetic spin models in two dimensions: the
$q=3$ states Potts model and the Ising model. It is known that, in
agreement with the Harris criterion\cite{HARRIS}, the disorder is
relevant for the $q=3$ states Potts model, i.e. the renormalization
group flows to a stable fixed point, which is different from the fixed
point of the pure system\cite{Ludwig,ddp}. For the Ising model the
disorder is marginally irrelevant and the renormalization group flows
back to the fixed point of the pure system with logarithmic
corrections\cite{dd}.

In the case of the Potts model we find very similar behavior as in the
RFIM. A scaling analysis shows that, as in the RFIM, the violation of
self-averaging of the CL, although weaker than in the RFIM, persists
for arbitrarily large systems in the vicinity of the critical point.

In the case of the Ising model, the violation of self-averaging is
still weaker. The scaling analysis seems to work but we cannot exclude
a very slow restoration of self-averaging in the thermodynamic limit
(see later).

We conclude that the non self-averaging of the CL is not a singularity
of the RFIM at zero temperature, but a more general phenomenon in the
physics of disordered systems. The correlation length plays a crucial
role in the scaling theories of phase transitions. It is not clear yet
what are the consequences of the absence of self-averaging of the CL
for those scaling theories in disordered systems.

In both the Potts and the Ising models we consider the case of square
lattices of size $ L^2 $ where $L=200,\ 350 $, and $ 500 $, with
nearest neighbor interactions. The ferromagnetic couplings $J $ are
independent random variables which can take the two values $J = J_1 $
and $J = J_2 $ with equal probability. It is well known that these
models are self-dual and that the critical inverse temperature $ \beta
= \beta_c $ is given by \cite{Kinzel}:
\begin{equation} 
\label{tc}
 (\exp{\beta_{c} J_1} -1 ) ( \exp{\beta_{c} J_2} -1 ) = \ \ q \; .
\end{equation}
The knowledge of $ \beta_c $ greatly simplifies the scaling analysis.
It is also known that cluster algorithms\cite{sww} are very effective
for these models. We simulated 1000 samples for $L=500$, $2000 $ for
$L=350$ and $10000$ for $L=200$. As in the case of the RFIM\cite{PS0},
we measured the correlation length by studying the dependence on the
boundary conditions.  More precisely in the Ising case (the
generalization to the Potts case is obvious) we set all the spins
equal to one on the line $x=0$ and choose free boundary conditions at
the other end of the lattice $ x = L $.  We impose periodic boundary
conditions on the perpendicular direction.  After thermalisation, for
every sample $s $ we measure the magnetization $ m_s (x ) = 1/L
\sum_{y} < \sigma(x,y) > $ and study its dependence on $x$ and on the
inverse temperature $ \beta $.  For the Potts model we studied
extensively the case $ J_1 / J_2 = 10$\cite{jacpic} and $ \beta /
\beta_c = .95,\ .96,\ .97,\ .98,\ .99,\ .995 $ and $ 1 $. We verify
thermalisation in the following ways. For every realization of the
couplings, we simulate two copies with different initial conditions.
In one copy all spins are set to $ + 1 $ and in the other they are
chosen at random (except on the line $ x = 1 $, where they are set to
$ + 1 $). Thermalisation time $ t_{th} $ must be long enough in order
for the magnetization $ m_s(x) $ of the 2 copies to be
indistinguishable. The local magnetizations $ < \sigma(x,y) > $ are
the time averages of the local spins during the measuring time $ t_m $
We measure the autocorrelation time $ \tau $ and require both the
thermalisation time $ t_{th} $ and the measuring time $ t_m $ to be
much larger than $ \tau $. We also verify that there are no time
drifts in the measurements, i.e. averages over the first half of $ t_m
$ do not differ from measurements over the second half of $ t_m $.
Even at $ \beta = \beta_c $ and for the volumes we simulated, $ t_{th}
$ is finite and reasonably small because of the cluster algorithm. For
$q=3$, we measured $\tau_{L=200}\simeq 200, \tau_{L=350}\simeq 400,
\tau_{L=500}\simeq 800$ and we choose $t_{th}=10 \tau, t_m = 20 \tau$.

To reduce errors, we found convenient to use a variance reduction
technique: we measured the probability $ p(x,y) $ of the spin $
\sigma(x,y) $ to be on the same cluster with the spins at the boundary
$ x=0 $. Obviously $ < \sigma(x,y) > = 2 p(x,y) - 1 $. In this way we
eliminate statistical fluctuations due to flips of large clusters.

In order to study sample to sample fluctuations, we measured, as in
the case of the RFIM\cite{PS0}, $ m(x) = \overline { m_s(x) } $ and $
m_{(2)} (x) = \overline { m_s(x)^2 } $, where, as usually, the bar
denotes average over the random coupling samples.  As it was mentioned
earlier, for every sample we simulated two copies, differing only at
the initial conditions.  In order to reduce errors, we measure the
probability $ p_2 (x,y) $ that in both of the two copies we simulated,
the spin at the site $ (x,y) $ belongs to the cluster connected to the
boundary at $ x=0 $.  Obviously $ < \sigma_1 (x,y) > < \sigma_2 (x,y)
> = 2 p_2 (x,y) - 1 $ where $ \sigma_1 $ and $ \sigma_2 $ are the
spins of the two copies. With these definitions, $ m_{(2)} (x) = 1 / L
\sum_{y} 2 \overline{p_2 (x,y) } - 1 $.

For $ \beta < \beta_c $, we expect $ m(x) $ and $ m_{(2)} (x) $ to be
exponentially decreasing functions of $x$.
\begin{figure}[t]
\rotatebox{270}
{\epsfxsize=10cm\epsffile{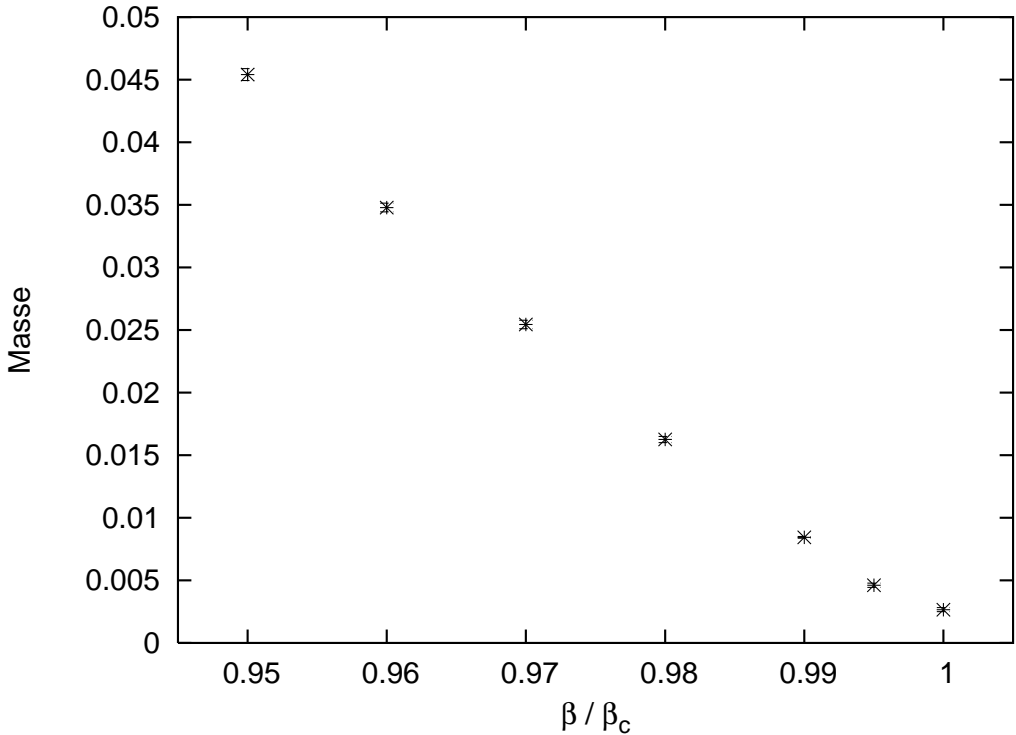}}
\smallskip
\caption{Mass as a function of $ \beta < \beta_c $ for the
$q=3$ Potts model and $L=500 $.
\label{fig1}}
\end{figure}
We  fitted $ m(x) $ with the formula
\begin{equation}
 m (x) \ = \ { \exp{( -\mu x )} \over x^{\alpha } } (a + { b
 \over x } ) \; .
\label{eq2}
\end{equation}
We found that this formula fits very well the data for all values of
$x$ except near the boundaries $ x \sim 0 $ and $ x \sim L $ and this
for all values of $ \beta $ we studied. The mass $ \mu $ (or inverse
correlation length) for the $q=3$ states Potts model, $ J_1 / J_2 = 10
$ and $ L= 500 $ is plotted in figure \ref{fig1} as a function of $
\beta < \beta_c $.  Throughout this paper we used the jack-knife
method to estimate statistical errors. We found that statistical
errors are smaller than systematic errors. We applied a small $ x $
cutoff $ x_0 $ and a large $ x $ cutoff $ x_1 $ in fitting the data,
because, as expected, close to the boundaries the previous exponential
form is not valid. The choice of $ x_0 $ and $ x_1 $ is an important
source of systematic errors on the mass $ \mu $. For every size $L$
and every $ \beta $ we tried all pairs of $ x_0 $ and $ x_1 $ for
which the fit was reasonable, i.e. the $ \chi^2 $ of the fit per
degree of freedom was $ .10 \le \chi^2 \le .80 $ (the values of $ m(x)
$ at different $x$ are not statistically independent). The error bars
in figure \ref{fig1} reflect the dependence of the mass $ \mu $ on the
different choices of ``acceptable'' cutoff's (statistical errors are
much smaller). Typical values were $ 5 \le x_0 \le 15 $ and $ .5 L \le
x_1 \le .8 L $ depending on $L$ and $\beta $.  We performed a similar
analysis of $ m_{(2)} (x) $ defined above and measured in the same way
the corresponding mass $ \mu_2 $. The ratio $ R = \mu_2 / \mu $ is
dimensionless. It is plotted in figure \ref{fig2} as a function of the
scaling variable $ L^{1/ \nu } ( \beta_c - \beta ) / \beta_c $ for the
$q=3$ states Potts model and $ J_1 / J_2 = 10 $.  Following the
results of previous simulations \cite{jc}, we assumed that the
critical exponent $\nu =1$.

\begin{figure}[t]
\rotatebox{270}
{\epsfxsize=10cm\epsffile{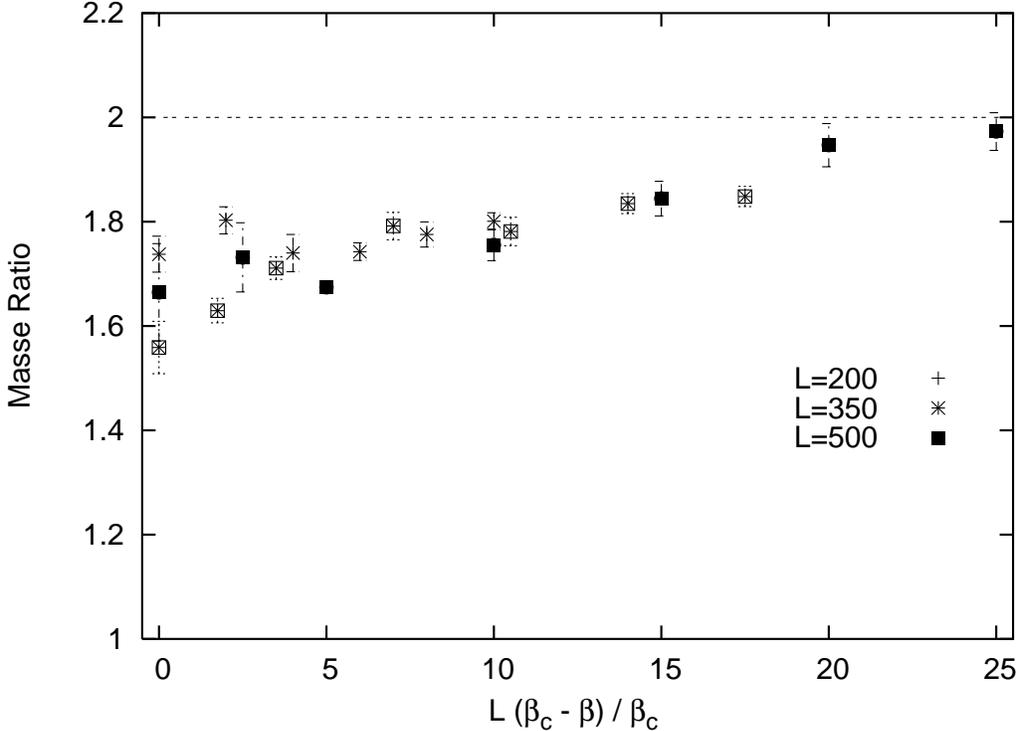}}
\smallskip
\caption{Finite size scaling plot of the mass ratio (see text)
versus  $ L^{1/ \nu } ( \beta_c - \beta ) / \beta_c $ for the
$q=3$ Potts model.
\label{fig2}}
\end{figure}

Figure \ref{fig2} shows that the behavior of $ R $ is compatible with
the scaling ansatz. For $ \beta $ far from $ \beta_c $ $ R=2 $,
i.e. for large $ x $, $ m_{(2)} (x) $ decays twice as fast as $ m (x)
$.  This shows that, far from criticality, the correlation length is
self-averaging.

Near the critical point, $ R < 2 $. As this is discussed in \cite{PS0}
this implies the existence of a bound state in the underlying field
theory. In the case of the RFIM $ R = 1 $, which means that the bound
state completely dominates the asymptotic long distance behavior. In
the $ q=3 $ random Potts model $ 1 < R < 2 $, which suggests a weaker
coupling of the bound state, as we will see.

As this is discussed in \cite{PS0} if there exist more than one
comparable mass scales, as it can be in the case of bound states, we
expect a superposition of exponentials governing the long distance
behavior. This was the case for the RFIM.

We performed a similar analysis and fitted our data as follows:
\begin{equation} 
\label{d2}
m_{(2)}(x) \ = \ c_1 m (x)  +   c_2 ( m (x))^2 \; .
\end{equation}
\begin{figure}[t]
\rotatebox{270}
{\epsfxsize=10cm\epsffile{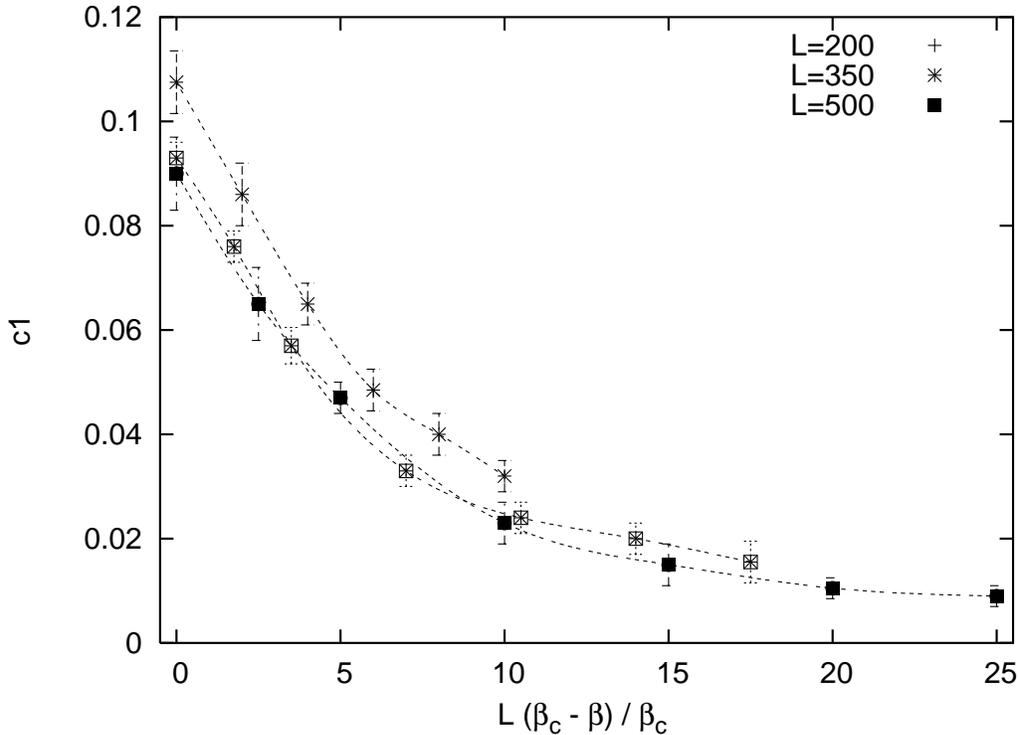}}
\smallskip
\caption{Finite size scaling plot of $ c_1 $ (see text) for the
$q=3$ Potts model.
\label{fig3}}
\end{figure}
It turned out that this two parameter fit of $ m_2(x) $ is better than
the four parameter fit of eq.~(\ref{eq2}). If $ c_1 \ne 0 $ the first
term dominates for large $x$ ($ m (x) $ decreases much slower than $ m
(x)^2 $) and the mass is non self-averaging.  As for the RFIM we made
the finite size scaling hypothesis that the dimensionless coefficient
$ c_1 $ is a function of $ L^{1/ \nu } ( \beta_c - \beta ) / \beta_c
$.  In figure \ref{fig3} we plot $ c_1 $ versus $ L ( \beta_c - \beta
) / \beta_c $ (we took $ \nu = 1 $) for different values of $ \beta $
and $ L = 200, \ 350 $ and $500 $.  We see that for $L=500$ and
$L=350$ the data fall completely on each other, while for $L=200$
small non leading corrections to scaling seem to be present.  For $
\beta = \beta_c $, $c_1 = .10 \pm .02$. We conclude that for the $ q=3
$ random Potts model the mass is not self-averaging for any finite
volume, provided the temperature is close enough to its critical
value. This is due to the presence of bound states in the replica
field theory, as in the case of the RFIM.  The violation of
self-averaging is much weaker than for the random field model in three
dimensions where we found $c_1 = 1 $, i.e the maximum possible
violation of self-averaging.

\begin{figure}[t]
\rotatebox{270}
{\epsfxsize=10cm\epsffile{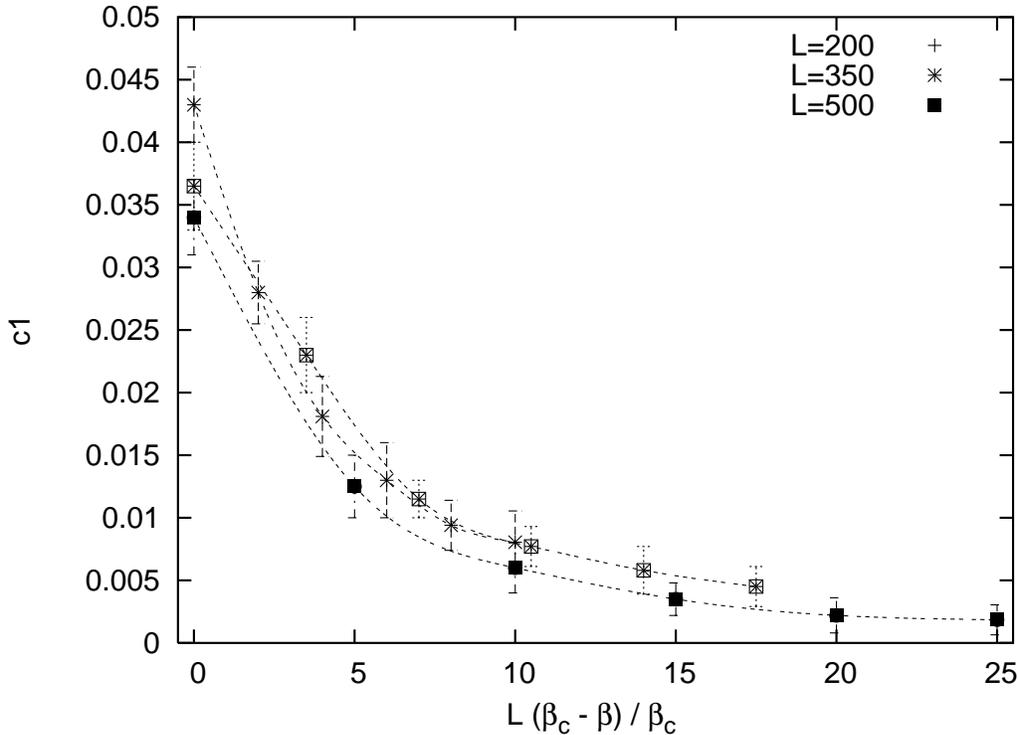}}
\smallskip
\caption{Finite size scaling plot of $ c_1 $ (see text) for the
Ising model.
\label{fig4}}
\end{figure}

We performed a similar analysis for the Ising model with $ J_1 / J_2 =
5 $.  In figure \ref{fig4} we plot $ c_1 $ versus $ L ( \beta_c -
\beta ) / \beta_c $.
At first sight the data seem again compatible with this scaling
hypothesis and there is again evidence of violation of self-averaging
of the mass, although weaker than in the Potts case ( $ c_1 \sim .033
$ at $ \beta_c $). A closer look to the data at $ \beta = \beta_c $,
however, leaves open the possibility of a slow crossover to $ c_1 = 0
$ at $ \beta_c $ as $ L \to \infty $. (Away from $ \beta_c $ scaling
seems to work fine.)  In order to settle this question we simulated a
larger lattice with $ L = 700 $ at $ \beta_c $. If scaling of $ c_1 $
holds, $ c_1 $ should be independent of $ L $ at $\beta = \beta_c $.
We found the following values of $ c_1 $ at $ \beta_c $: { $ .034 \pm
.003 $ for $L=700$, $ .0355 \pm .004 $ for $L=500 $, $ .0365 \pm .004
$ for $ L=350 $ and $ .0425 \pm .004 $ for $ L = 200 $}. $ c_1 $ seems
to decrease from $L=200$ to $L=350 $ and then change very little from
$L=350 $ to $L=500 $ and to $L=700 $.  It is known\cite{dd} that the
renormalization group fixed point of the random Ising model flows to
the pure Ising fixed point at a logarithmic rate. Our data cannot
discriminate between such a behavior or sub-dominant corrections to
scaling when the lattice is not large enough.

Finally in order to verify the dependency of the violations of
self-averaging of the CL on the disorder, i.e. the ratio $ r = J_1 /
J_2 $, and the number of states of the Potts model $ q $, we measured
$c_1$ for $ L=200 $ and $\beta =\beta_c $ for other values of $q $ and
$r$.  For $q=8$ and $r=10$ we got $c_1 = .19 \pm.02 $. For $q=2$
(Ising) we got $c_1 = .0145 \pm.005 $ for $ r = 2 $ and $c_1 = .063
\pm.002 $ for $ r = 10 $. We conclude that $c_1$, i.e. the violations
of self-averaging of the correlation length, increase when the
disorder is larger, or when the number $q$ of states of the Potts
model increases.

\end{document}